\documentclass[aps,twocolumn,nofootinbib,showpacs]{revtex4}
\usepackage{bm,amssymb,graphicx}
\usepackage[colorlinks,linkcolor=magenta,urlcolor=cyan,citecolor=blue]{hyperref}

\begin{document}
\bibliographystyle{apsrev}

\title{An Alternative to the Lagrangian and Hamiltonian Formulations of Relativistic Field Theories Based on the Energy-Momentum Tensor}

\author{Hans Christian \"Ottinger}
\email[]{hco@mat.ethz.ch}
\homepage[]{http://www.polyphys.mat.ethz.ch}
\affiliation{ETH Z\"urich, Department of Materials, Polymer Physics, HCP F 47.2,
CH-8093 Z\"urich, Switzerland}

\date{\today}

\begin{abstract}
A Noether-enhanced Legendre transformation from Lagrange densities to energy-momentum tensors is developed into an alternative framework for formulating classical field equations. This approach offers direct access to the Hamiltonian while keeping manifest Lorentz covariance in the formulation of relativistic field theories. The field equations are obtained by imposing a vanishing divergence of the energy-momentum tensor (in a suitably structured form). The proposed framework is ideally suited for coupling subsystems because their interaction can be expressed as an exchange of energy and momentum. Even higher derivative theories and dissipative systems can be treated. A most promising application is the formulation of alternative theories of gravity. The proposed framework is illustrated for Yang-Mills theories, for which it offers a covariant canonical quantization scheme.
\end{abstract}

\pacs{03.50.-z, 03.65.Ca, 11.10.Ef}

% 03.50.-z Classical field theories
% 03.65.Ca Formalism
% 03.70.+k Theory of quantized fields
% 11.10.Ef Lagrangian and Hamiltonian approach

\maketitle

\section{Motivation}
Problems in classical mechanics and field theory are usually treated within the Hamiltonian or Lagrangian frameworks, where the two are connected through Legendre transformation. The respective advantages of the one or other approach depend on the specific problems and interests. The Hamiltonian approach has the advantage of a nicely structured underlying phase space offering a geometric interpretation, whereas symmetries can be treated more elegantly in the Lagrangian approach. Arguably, the Hamiltonian approach offers a more robust quantization procedure. Our goal here is to develop an alternative framework that gives us direct access to the Hamiltonian (and further observable quantities) but nevertheless keeps the manifest Lorentz covariance of relativistic field theories.

The proposed approach is based on an enhanced Legendre transformation from Lagrange densities to energy-momentum tensors, so that the standard transformation from Lagrange to Hamiltonian densities appears as one of the tensor components. The enhanced transformation is expected to provide major advantages, for example, in coupling subsystems through an exchange of energy and momentum, in describing dissipative systems, and in developing alternative theories of gravity, including higher derivative theories. For the example of Yang-Mills theories, we illustrate how constraints can be handled and how a covariant canonical quantization scheme can be obtained.

\section{Framework}
We consider a set of fields $\varphi^a$ on an underlying Minkowski space, labeled by a discrete superscript $a$. A field theory be specified by a Lorentz invariant Lagrangian density ${\cal L}(\varphi^a, \varphi^a_{,\mu})$, where the subscript ``$,\mu$'' on a field indicates the partial derivative with respect to the space-time coordinates $x^\mu$ of a fixed reference frame (similarly, a superscript ``$,\mu$'' indicates a derivative with respect to $x_\mu$). The space-time integral of the Lagrangian density defines the action
\begin{equation}\label{action}
   I = \frac{1}{c} \int {\cal L}(\varphi^a, \varphi^a_{,\mu}) \, d^4x .
\end{equation}
In the Lagrangian approach, stationarity of the action functional leads to the typically second-order field equations for the fields $\varphi^a$.

Our further discussion is based on the tensor
\begin{equation}\label{enhancedLegendre}
   {T_\mu}^\nu = {\delta_\mu}^\nu {\cal L}
   - \varphi^a_{,\mu} \, \frac{\partial{\cal L}}{\partial\varphi^a_{,\nu}} ,
\end{equation}
where ${\delta_\mu}^\nu$ is the Kronecker delta. For $\mu = \nu = 0$, we recognize a Legendre transformation in this definition. The sign convention is chosen such that $T^{00} = - {T_0}^0 = {\cal H}$ is the Hamiltonian density, that is, it depends on our choice for the signature of the Minkowski metric $(-+++)$. The Hamiltonian density ${\cal H}$ is not a scalar density but rather a component of a tensor, which turns out to be (related to) the energy-momentum tensor. This lack of Lorentz invariance of the Hamiltonian density is a consequence of focussing on time evolution in the Hamiltonian approach. We here treat time and space, or energy and momentum, on a more equal footing in order to keep Lorentz invariance.

Why should it be physically meaningful to transform from a scalar to an entire tensor? The answer to this question rests in the identity
\begin{equation}\label{localconservation}
   \frac{\partial {T_\mu}^\nu}{\partial x^\nu} =
   c \, \varphi^a_{,\mu} \, \frac{\delta I}{\delta\varphi^a} ,
\end{equation}
which is obtained by calculating the derivative $\partial {\cal L} / \partial x^\mu$ and assuming that ${\cal L}$ depends on $x^\mu$ only through the fields $\varphi^a$ and $\varphi^a_{,\mu}$, but not explicitly (see Appendix~\ref{appenymomtens} for a detailed derivation). This assumption expresses the invariance of a field theory under space-time translations. According to Noether's theorem, momentum and energy must be conserved or, for a field theory, their densities must be divergence free. Indeed, the Euler-Lagrange equations resulting from the stationarity of the action (\ref{action}) imply that the right-hand side of Eq.~(\ref{localconservation}) vanishes. As the physical significance of the transformation from scalar to tensor in Eq.~(\ref{enhancedLegendre}) stems from the local conservation of energy and momentum, we refer to it as a Noether-enhanced Legendre transformation.

An alternative framework for formulating relativistic field theories, independent of the existence of an underlying action, can now be formulated as follows: The fundamental ingredient is an energy-momentum tensor ${T_\mu}^\nu$ depending on the basic fields $\varphi^a$ (and their space-time derivatives) with the property
\begin{equation}\label{localconservationform}
   \frac{\partial {T_\mu}^\nu}{\partial x^\nu} =
   \varphi^a_{,\mu} \, \eta_a ,
\end{equation}
where the auxiliary fields $\eta_a$ depend on $\varphi^a$, $\varphi^a_{,\mu}$, and $\varphi^a_{,\mu,\nu}$. In terms of these auxiliary fields, the field equations are then obtained as
\begin{equation}\label{fieldequationform}
   \eta_a = 0 .
\end{equation}
The field equations are contained in the requirement that the energy-momentum tensor must be divergence-free to express the local conservation of energy and momentum. For this reason one could refer to the proposed framework as the transport perspective on field theories. Through the assumed structure of the right-hand side of Eq.~(\ref{localconservationform}) the framework can, in general, provide more than the four equations describing the transport of energy and momentum. All classical field theories involve the transport of energy and momentum. After quantization, energy and momentum are carried by the field quanta.

\section{Transformations}
The tensor ${T_\mu}^\nu$ is not unique because one can add any divergence-free tensor. For example, a transformation considered by Belinfante \cite{Belinfante40},
\begin{equation}\label{transformation}
   {T_\mu}^\nu \rightarrow {T_\mu}^\nu + \frac{\partial}{\partial x^{\nu'}}
   ( {K_\mu}^{\nu\nu'} - {K_\mu}^{\nu'\nu}) ,
\end{equation}
leaves Eq.~(\ref{localconservationform}) invariant for any choice of the tensor ${K_\mu}^{\nu\nu'}$. Whereas the tensor ${T_\mu}^\nu$ defined in Eq.~(\ref{enhancedLegendre}) is often referred to as the energy-momentum tensor (see, for example, Eq.~(11.48) of \cite{BjorkenDrell}), one may prefer a more physical, symmetric energy-momentum tensor that can be achieved by a transformation of the type (\ref{transformation}). A more systematic procedure leading to symmetry has actually been developed in \cite{MontesinosFlores06}. One should, however, keep in mind that the proposed framework for formulating relativistic field theories works only if the transformation of the energy-momentum tensor is achieved without using the field equations, which are supposed to be the output of the framework. We next illustrate this important point by means of a relevant example.

\section{Yang-Mills theories}
For illustrating the general framework based on energy-momentum tensors, we consider the example of Yang-Mills theories. Of course, these theories are of eminent importance in the modern theory of fundamental particles.

\subsection{Energy-momentum tensor}
We consider a set of fields $A_{a \mu}$ labeled by two indices: a label $a$ for the base vectors of a Lie algebra (associated with a continuous symmetry group) and the space-time index $\mu$, which makes $A_{a \mu}$ a four-vector field for each $a$. We use exactly the same notation as in \cite{hco229}, except that we distinguish between upper and lower Lie algebra labels $a$ (which is unnecessary for the usual cases of the special unitary groups ${\rm SU}(2)$ and ${\rm SU}(3)$ associated with weak and strong interactions, but does matter for the restricted Lorentz group ${\rm SO}(1,3)$). In terms of the field tensors
\begin{equation}\label{YMFdefinition}
   F_{a \mu\nu} = A_{a \nu ,\mu} - A_{a \mu ,\nu}
   - g f^{bc}_a A_{b \mu} A_{c \nu} ,
\end{equation}
where the parameter $g$ characterizes the strength of the interaction and the quantities $f^{bc}_a$ are the structure constants of the Lie algebra, the Lagrangian density of Yang-Mills theories \cite{YangMills54,PeskinSchroeder,WeinbergQFT2} can be written as
\begin{equation}\label{YML}
   {\cal L} = - \frac{1}{4} F_{a\mu\nu}  F^{a\mu\nu} ,
\end{equation}
where the index $a$ is raised (or lowered) by the Cartan-Killing metric of the Lie algebra in the same way as space-time indices are raised (or lowered) by the Minkowski metric. By means of Eqs.~(\ref{enhancedLegendre}) and (\ref{transformation}) we can construct the energy-momentum tensor
\begin{equation}\label{YMenergmom}
   {T_\mu}^\nu = {\delta_\mu}^\nu {\cal L}
   + F_{a \mu\rho} F^{a\nu\rho} + A_{a \mu} \left( \frac{\partial F^{a\rho\nu}}{\partial x^\rho}
   - g f^{ab}_c A_{b \rho} F^{c\rho\nu} \right) .
\end{equation}
Note that ${T_\mu}^\nu$ contains second derivatives of the basic fields $A_{a \mu}$, which result from the transformation (\ref{transformation}). As a consequence of the antisymmetry of $F^{a\nu\rho}$, the divergence of ${T_\mu}^\nu$ does, however, not contain any third derivatives. By comparing the divergence of the tensor in Eq.~(\ref{YMenergmom}) to Eq.~(\ref{localconservationform}), we get the auxiliary fields
\begin{equation}\label{YMauxiliaryfields}
   \eta^{a \nu} = \frac{\partial F^{a\mu\nu}}{\partial x^\mu}
   - g f^{ab}_c A_{b \mu} F^{c\mu\nu} ,
\end{equation}
and the second-order Yang-Mills field equations in the form $\eta^{a \nu} = 0$. From the simpler energy-momentum tensor ${\delta_\mu}^\nu {\cal L} + F_{a \mu\rho} F^{a\nu\rho}$, which is obtained from Eq.~(\ref{YMenergmom}) by making use of the field equations, a more complicated divergence condition would arise (actually, a higher-order differential equation) and a straightforward identification of the auxiliary fields would no longer be possible.

\subsection{Canonization}
With the goal of quantization in mind, we would like to recognize a canonical structure in Yang-Mills theories. Using the standard procedure for constructing the conjugate momenta associated with the components of the four-vector potential $A_a^\mu$, we consider
\begin{equation}\label{conjugatemomentastand}
   \frac{\partial{\cal L}}{\partial A^\mu_{a \,,0}} = F^a_{0\mu} .
\end{equation}
By a comparison to the field tensor of electrodynamics, the spatial components of $F^a_{0\mu}$ can be recognized as electric-field-like variables (except for a sign to be changed below). However, as a consequence of the antisymmetry $F^a_{\nu\mu}=-F^a_{\mu\nu}$, we have $F^a_{00}=0$, so that there is no conjugate momentum associated with $A_a^0$.

A possible strategy would be to eliminate $A_a^0$ (and possibly also further variables) to be left with conjugate pairs of variables. However, we strongly prefer to keep the full four-vectors $A_a^\mu$ in order to provide a Lorentz covariant setting. Therefore, we introduce the additional Lorentz scalar fields $E^a$ into Yang-Mills theories which, in our fixed reference frame, can serve as the conjugate momenta associated with the basic fields $A_a^0$. We introduce these additional variables by modifying the energy-momentum tensor (\ref{YMauxiliaryfields}). Of course, this modification cannot be of the type (\ref{transformation}) because we need additional equations for the new variables and moreover expect a modification of the other field equations by the new fields. We propose the following addition to the energy-momentum tensor (\ref{YMenergmom}),
\begin{equation}\label{spaceexpansion1}
   {T_\mu}^\nu \rightarrow {T_\mu}^\nu + \frac{1}{2} {\delta_\mu}^\nu \, E^a E_a
   + E^a ( A^\nu_{a ,\mu} - {\delta_\mu}^\nu A^\rho_{a ,\rho} ) .
\end{equation}
For this transformation, we do not need to insist on an underlying variational principle, but we must insist on the structure (\ref{localconservationform}) of the divergence. Our guiding principles are the proper introduction of the extra field variables $E^a$ and a minimal perturbation effect on the original field equations. Indeed, we find
\begin{equation}\label{spaceexpansion2}
   \frac{\partial {T_\mu}^\nu}{\partial x^\nu} \rightarrow
   \frac{\partial {T_\mu}^\nu}{\partial x^\nu} + A_{a \nu \,,\mu} \, E^{a \,,\nu}
   + E^a_{\,,\mu} \, ( E_a - A^\nu_{a \,,\nu}) ,
\end{equation}
which implies a modification of the auxiliary fields (\ref{YMauxiliaryfields}) and of the corresponding field equations according to
\begin{equation}\label{YMauxiliaryfieldsmod}
   \eta^{a \nu} \rightarrow \eta^{a \nu} + E^{a \,,\nu} = 0 ,
\end{equation}
and the new field equations
\begin{equation}\label{YMaddeq}
   E_a - A^\nu_{a \,,\nu} = 0 .
\end{equation}
The latter equations may be considered either as a definition of the new variables $E_a$ or as (the previously missing) time evolution equations for the fields $A^0_a$.

We are now in a position to identify a useful structure in the enlarged space. For that purpose, we define the four-vector field
\begin{equation}\label{E4def}
   E^a_\mu = F^a_{\mu\nu} u^\nu - E^a u_\mu ,
\end{equation}
where, in the underlying fixed reference frame, we define $u^\nu=(1,0,0,0)$. These electric-field-like variables $E^a_\mu$ are the conjugate partners of the vector potentials $A_a^\mu$ of our Yang-Mills theory. We expect Lorentz invariant canonical Poisson brackets for the conjugate fields $E^a_\mu$ and $A_a^\mu$, which could serve as a starting point for quantization. To verify that $E^a$ and $A^0_a$ are canonical conjugates we consider the identity
\begin{equation}\label{conjugatescheck}
   A^0_{a ,0} = - \frac{\partial {\cal H}}{\partial E^a} =
   \frac{\partial {T_0}^0}{\partial E^a} = E_a - A^j_{a ,j} ,
\end{equation}
which coincides with the new field equations (\ref{YMaddeq}). The construction implied by Eqs.~(\ref{spaceexpansion1}) and (\ref{E4def}) is perfectly consistent with the Hamiltonian approach to the quantization of Yang-Mills theories on Fock space in \cite{hco229}.

If we reconstruct the modified Lagrangian density from the Hamiltonian density ${\cal H}$, the result
\begin{equation}\label{Lexpansion}
   {\cal L} \rightarrow {\cal L} - \frac{1}{2}  A^\mu_{a ,\mu} A^{a \nu}_{,\nu} ,
\end{equation}
implies the following modification of the energy-momentum tensor (\ref{YMenergmom}),
\begin{equation}\label{LexpansionT}
   {T_\mu}^\nu \rightarrow {T_\mu}^\nu
   - \frac{1}{2} A^{\mu'}_{a ,\mu'} A^{a \nu'}_{,\nu'} \, {\delta_\mu}^\nu
   + A^{a \nu'}_{,\nu'} \, A^\nu_{a,\mu} .
\end{equation}
The two modifications (\ref{spaceexpansion1}) and (\ref{LexpansionT}) coincide only if the field equation (\ref{YMaddeq}) is used. The modification of the auxiliary fields (\ref{YMauxiliaryfields}) is now obtained as
\begin{equation}\label{YMauxiliaryfieldsmodx}
   \eta^{a \nu} \rightarrow \eta^{a \nu} + (A^{a \mu}_{,\mu})^{,\nu} = 0 ,
\end{equation}
which is a combination of Eqs.~(\ref{YMauxiliaryfieldsmod}) and (\ref{YMaddeq}). In short, it would seem to be natural to replace the Yang-Mills Lagrangian density (\ref{YML}) by the canonical version
\begin{equation}\label{YMLcanonical}
   {\cal L} = - \frac{1}{4} F_{a\mu\nu}  F^{a\mu\nu}
   - \frac{1}{2}  A^\mu_{a ,\mu} A^{a \nu}_{,\nu} ,
\end{equation}
which is still Lorentz invariant, but no longer gauge invariant. Therefore, this canonical version of Yang-Mills theories should eventually be supplemented by the Lorenz gauge $A^\mu_{a ,\mu}=0$.

\subsection{Momentum density}
The components ${T_\mu}^0$ can be interpreted as the spatial densities of energy and momentum. One usually focuses on the Hamiltonian density ${\cal H}$ because the Hamiltonian generates time evolution. We here would like to pay attention also to the momentum density $\bm{{\cal M}}$ of Yang-Mills theories. Equations (\ref{YMenergmom}) and (\ref{spaceexpansion1}) imply the remarkably simple result
\begin{equation}\label{YMmomentumdensity}
   \bm{{\cal M}} = E^a_\mu \bm{\nabla} A_a^\mu - \bm{\nabla} \cdot ( \bm{E}^a \bm{A}_a ) .
\end{equation}
As the divergence term does not contribute to the total momentum obtained by integrating the density $\bm{{\cal M}}$ over the entire space, one can easily recognize that the total momentum generates infinitesimal space translations of the canonical fields.

\subsection{Comments on quantization}
As we have identified $E^a_\mu$ and $A_{a \mu}$ as canonically conjugate fields, the quantization of the free Yang-Mills fields is obtained by postulating the canonical quantization rules
\begin{equation}\label{YMcanoquantization}
   \left[ E^a_\mu(\bm{x},t) , A_b^\nu(\bm{x}',t) \right] =
   i \, {\delta_\mu}^\nu {\delta^a}_b \, \delta^3(\bm{x}-\bm{x}') .
\end{equation}
Further commutators can be inferred from these relations by taking spatial derivatives (however, inconsistencies may arise from the inherent subtleties of quantum field theories as is well-known, for example, from the problem of the Schwinger term \cite{Schwinger59,NishijimaSasaki75,Kubo94,Kinoshita}).

The only nonvanishing commutators between the canonical vector variables and the field tensor components are given by
\begin{equation}\label{YMcommutatorFA}
   \left[ F^a_{j0}(\bm{x},t) , A_b^k(\bm{x}',t) \right] =
   i \, {\delta_j}^k {\delta^a}_b \, \delta^3(\bm{x}-\bm{x}') ,
\end{equation}
and
\begin{eqnarray}
   \left[ E^a_j(\bm{x},t) , F_c^{jk}(\bm{x}',t) \right] &=&
   - \left[ E^a_j(\bm{x},t) , F_c^{kj}(\bm{x}',t) \right] =
   \nonumber \\ && \hspace{-8em}
   i \left( {\delta^a}_c \frac{\partial}{\partial x_k} - g f^{ab}_c A_b^k \right)
   \delta^3(\bm{x}-\bm{x}') ,
\label{YMcommutatorEF}
\end{eqnarray}
for $j \neq k$ (no summation over $j$). Two components of the field tensor commute unless one index pair is of the type $0j$ (or $j0$) and the other pair is of the type $jk$ (or $kj$) with $j \neq k$. In view of the identity $F^a_{j0}=E^a_j$, all nonvanishing commutators among the components of the field tensor can actually be obtained from Eq.~(\ref{YMcommutatorEF}). Instead of deducing formal commutators from the canonical commutation relations (\ref{YMcanoquantization}), one could use them to construct a carefully regularized Fock space \cite{hco229,hcoqft}. Although the canonical commutation relations (\ref{YMcanoquantization}) are formulated for Lorentz four-vectors, they are not yet covariant because they involve equal-time commutators and a three-dimensional Dirac $\delta$ function. Instead of separating space and time in the underlying reference frame, we could replace the equal-time condition by $u_\mu(x^\mu-x'^\mu)$ and localize the $\delta$ function on the corresponding subspace.

The canonical quantization (\ref{YMcanoquantization}) is achieved in an unphysically large space because no gauge conditions have been fixed yet. In the well-known BRST approach (where the acronym BRST refers to the original papers by Becchi, Rouet, Stora \cite{BecchiRouetStora76} and by Tyutin \cite{Tyutin75}; for a pedagogical BRST primer, see \cite{Nemeschanskyetal86}), one first introduces an even larger space. The fields characterizing gauge transformations are introduced as additional ghost fields (with canonical structure). The physical states are then selected by specifying BRST charges, where the gauge invariance of a theory corresponds to field equations conserving these BRST charges.

\subsection{Particle in electromagnetic field} We illustrate the coupling between a particle and a field for the motion of a charged particle in an electromagnetic field. In this special case of a Yang-Mills theory with a one-dimensional Lie algebra, the label $a$ and the structure constants can be dropped. In the Lagrangian approach, the coupling is obtained by adding the action
\begin{equation}\label{EMparticlecoupling}
   I_{\rm particle} = \int \left[ - m c^2 \frac{d\tau}{dt}
   + q \, \frac{dx^\mu(t)}{dt} A_\mu \big( x(t) \big) \right] dt ,
\end{equation}
to the field action (\ref{action}), where $m$ is the mass of the particle, $q$ its charge, $x^\mu(t)$ represents the particle trajectory, and $\tau$ its proper time defined by
\begin{equation}\label{propertimedef}
   \frac{d \tau}{d t} = \frac{1}{c}
   \sqrt{- \frac{d x^\mu(t)}{dt} \frac{d x_\mu(t)}{dt}} .
\end{equation}
The motion of the particle is characterized by the Euler-Lagrange equation
\begin{equation}\label{particleinEMfield}
   m \frac{d^2 x_\mu}{d \tau^2} = q F_{\mu\nu} \frac{d x^\nu}{d \tau} .
\end{equation}
In the present approach, the effect of charged matter on the electromagnetic field is obtained by simply adding the electric current density four-vector to the auxiliary field (\ref{YMauxiliaryfields}),
\begin{equation}\label{particleonEMfield}
   \eta^\nu \rightarrow \eta^\nu + \delta^3 \big(\bm{x}-\bm{x}(t)\big)
   \, q \frac{d x^\nu(t)}{d t} .
\end{equation}
The structure of the field equation requires that the divergence of the current density four-vector vanishes (see, e.g., Sect.~2.6 of \cite{Weinberg}). Adding divergence-free current density four-vectors to the auxiliary fields is the standard way of coupling Yang-Mills fields to matter.

\section{Coupling of fields} The proposed formulation of field equations based on the divergence of the energy-momentum tensor is ideally suited for coupling two fields because the right-hand side of Eq.~(\ref{localconservationform}) has the interpretation of a source or sink of energy and momentum. Any source term for one field must be canceled by a sink term for the other field, and vice versa. In the context of coupling electrodynamics and the hydrodynamics of fluids with charged components, this intuitive way of coupling the respective field equations by exchanging energy and momentum has already been illustrated in textbooks (see Sect.~2.8 of \cite{Weinberg} or Sect.~5.4 of \cite{hcomctp}).

The coupling via energy-momentum tensors even works if the field equations are not obtained from auxiliary fields as introduced in Eq.~(\ref{localconservationform}). In particular, one does not depend on an underlying variational principle. In the above-mentioned textbook examples, the field equations of hydrodynamics are even dissipative. One should note that even the variational formulation of perfect fluid dynamics is quite challenging \cite{Schutz70}, so that a non-variational framework is clearly desirable.

The vanishing divergence of the energy-momentum tensor provides four differential equations (in this paper, we always assume $3+1$ dimensions). Therefore, resolving the structure of the right-hand side of Eq.~(\ref{localconservationform}) is important only if the number of physical degrees of freedom exceeds four. For electrodynamics, there are only two physical degrees of freedom. For perfect fluid dynamics, we have five degrees of freedom, which is still tractable because we have an additional continuity equation expressing the mass balance (or, more generally, the conservation of some charge or a particle number). This is the reason why also dissipative fluids can be handled. The present framework is ideally suited for fluid dynamics because balance equations are at the heart of this field. For multi-component fluids, the number of degrees of freedom increases, however, one has the option of coupling the hydrodynamic equations for the individual species (through pressure, multi-component diffusion, chemical reactions, etc \cite{hcomctp}).

Also for a theory of gravity we expect the number of physical degrees of freedom to be smaller than four (general relativity has two physical degrees of freedom). As it has been argued that, from the geometric perspective on general relativity, the concept of energy-momentum tensor is of questionable value (because the energy of the gravitational field cannot be localized; see \S 20.4 of \cite{MisnerThorneWheeler} and also the discussion in \S~101 of \cite{LandauLifshitz2}), we need to consider a larger class of relativistic field theories before we can return to the peculiarities of gravity.

\section{Higher derivative theories}
In physics, we are usually interested only in second-order partial differential equations. There is a deeper reason for that: Higher derivative theories are prone to instability. However, in particular in the context of gravity and also of renormalization, there can nevertheless be good reasons for considering higher derivative theories.

\subsection{Energy-momentum tensor}
If the Lagrangian density depends also on second derivatives of the basic fields,
${\cal L} = {\cal L}(\varphi^a, \varphi^a_{,\mu}, \varphi^a_{,\mu ,\nu})$,
the Noether-enhanced Legendre transformation (\ref{enhancedLegendre}) needs to be further generalized as follows,
\begin{equation}\label{enhancedLegendre2}
   {T_\mu}^\nu = {\delta_\mu}^\nu {\cal L}
   - \varphi^a_{,\mu} \, \frac{\partial{\cal L}}{\partial\varphi^a_{,\nu}}
   - \left( \varphi^a_{,\mu,\nu'} - \varphi^a_{,\mu} \frac{\partial}{\partial x^{\nu'}} \right)
   \frac{\partial{\cal L}}{\partial\varphi^a_{,\nu,\nu'}} .
\end{equation}
The auxiliary fields $\eta_a$ are still given by $c \, \delta I/\delta\varphi^a$ (Eq.~(\ref{localconservation}) as the cornerstone of our development remains valid; see Appendix~\ref{appenymomtens} for a derivation), but they can now depend on up to fourth-order derivatives of $\varphi^a$. In other words, in general, the field equations (\ref{fieldequationform}) are fourth-order partial differential equations.

Whereas the further enhancement of the Legendre transformation in Eq.~(\ref{enhancedLegendre2}) is straightforward, higher derivative theories have a dangerous disposition to instability. The reasons for this instability become clear by passing from the Lagrangian to the Hamiltonian formulation of higher derivative theories, which has been developed in a classical paper by Ostrogradsky \cite{Ostrogradsky1850}. A detailed discussion can be found in an educational article by Woodard \cite{Woodard15}. Instabilities can be avoided by constraints \cite{Chenetal13} or complexification \cite{RaidalVeermae17}, where the physical relevance of such complex theories remains unclear. For the nested theories that are obtained by expressing the variables of standard Lagrangians in terms of more basic variables and their derivatives, one can identify the constraints required to ensure stability in a straightforward manner \cite{hco235}. This observation opens interesting possibilities.

\subsection{Field theoretic approach to gravity}
We here consider gravity as a field theory on flat Minkowski space (see, for example, \cite{Feynmanetal,Straumann00}). The basic field is the metric $g_{\mu\nu}$ and the Lagrangian density is given by ${\cal L} = -\sqrt{g} \, c^4 R/ (16\pi G)$, where $G$ is Newton's constant, $g$ is the absolute value of the determinant of $g_{\mu\nu}$, and $R$ is the curvature scalar, which involves first and second derivatives of $g_{\mu\nu}$ so that we are actually dealing with a higher derivative theory. The auxiliary fields are obtained from the standard result for the functional derivative of the action with respect to the metric (see, for example, Eq.~(12.4.3) of \cite{Weinberg}),
\begin{equation}\label{gravauxiliaryfields}
   \eta^{\mu\nu} = \frac{\sqrt{g} c^4}{16\pi G}
   \left( R^{\mu\nu} - \frac{1}{2} \bar{g}^{\mu\nu} R \right) ,
\end{equation}
where $R^{\mu\nu}$ is the Ricci tensor and $\bar{g}^{\mu\nu}$ is the inverse of the metric (note that this inverse is not obtained by raising the indices of $g_{\mu\nu}$ with the Minkowski metric). It is quite remarkable, that the resulting field equation $\eta^{\mu\nu} = 0$ is a second-order differential equation although the Lagrangian contains second derivatives. The corresponding energy-momentum tensor can be identified by means of Eq.~(\ref{localconservationform}). With the help of a Bianchi identity (see, for example, Eq.~(6.8.3) of \cite{Weinberg}), we obtain
\begin{equation}\label{gravenergmom}
   {T_\mu}^\nu = \frac{\sqrt{g} c^4}{8\pi G} g_{\mu\mu'}
   \left( R^{\mu'\nu} - \frac{1}{2} \bar{g}^{\mu'\nu} R \right) .
\end{equation}
The occurrence of the Einstein tensor in both the auxiliary fields and the energy-momentum tensor of the gravitational field is another remarkable feature of the field theoretic version of general relativity. Equation (\ref{gravenergmom}) is similar to Eq.~(7.6.4) of \cite{Weinberg}, but we do not exclude linear terms from the energy-momentum tensor. For a field theory of gravity in Minkowski space, we find a perfectly reasonable result for the energy-momentum tensor, which plays a key role in our general framework for relativistic field theories.

In the absence of an electromagnetic field, the action (\ref{EMparticlecoupling}) for the particle simplifies to
\begin{equation}\label{gravparticlecoupling}
   I_{\rm particle} =  - m c^2 \int \frac{d\tau}{dt} \, dt ,
\end{equation}
but gravity comes with a modified definition of proper time,
\begin{equation}\label{propertimedefg}
   \frac{d \tau}{d t} = \frac{1}{c}
   \sqrt{- g_{\mu\nu}\big(x(t)\big) \frac{d x^\mu(t)}{dt} \frac{d x^\nu(t)}{dt}} .
\end{equation}
This action characterizes geodesic motion,
\begin{equation}\label{geodesicparticlemotion}
   \frac{d^2 x^\mu}{d\tau^2} =
   - \Gamma^\mu_{\nu\nu'} \frac{d x^\nu}{d\tau} \frac{d x^{\nu'}}{d\tau} ,
\end{equation}
where the quantities $\Gamma^\mu_{\nu\nu'}$ are the Christoffel symbols representing the Levi-Civita connection associated with the the metric $g_{\mu\nu}$. The corresponding modification of the auxiliary fields,
\begin{equation}\label{particleongravfield}
   \eta^{\mu\nu} \rightarrow \eta^{\mu\nu} + \frac{1}{2} \delta^3 \big(\bm{x}-\bm{x}(t)\big)
   \, m \frac{d x^\mu(t)}{d t} \frac{d x^\nu(t)}{d t} \frac{dt}{d\tau} ,
\end{equation}
together with Eq.~(\ref{gravauxiliaryfields}), reproduces Einstein's field equation with the additional interpretation that the sum of the energy-momentum tensors for the gravitational field and for the particle is zero. The surprising result that the vanishing divergence of the stress tensor leads to a vanishing stress tensor is a consequence of a Bianchi identity, in combination with the presumed effect of gravity on proper time. From the perspective of the proposed approach, it is unnatural to obtain the field equations by equating energy-momentum tensors.

The coupling of the particle motion to the field for gravity is very different from the standard linear coupling for Yang-Mills fields employed in Eq.~(\ref{EMparticlecoupling}). It has been argued \cite{Feynmanetal,Deser70,Straumann00} that such a modification is required for consistency reasons. Under fairly general assumptions it turns out that any tensor theory of gravity must be equivalent to general relativity and that the underlying Minkowski space of the field theoretic approach is unobservable \cite{Straumann00} (for serious counter-arguments see, however, \cite{Padmanabhan08}). In any case, the field theoretic approach is useful by offering standard tools and for exploring possible alternatives to general relativity. According to Padmanabhan, ``There is sufficient evidence to assume that gravity is not a fundamental field but an emergent phenomenon like elasticity'' (see p.\,389 of \cite{Padmanabhan08}). In a much wider context, the possibility of emergent gravity has been discussed in the monograph \cite{Crowther}.

Is there any alternative theory of gravity for which the action, like for general relativity, depends on second derivatives of the basic fields but, unlike for general relativity, the field equations are of higher order (the natural expectation would be fourth-order field equations)? Based on the decomposition $g_{\mu\nu} = {b_\mu}^\kappa \, b_{\kappa\nu}$, one can introduce gauge vector fields $A_{a \mu}$ in an Ashtekar-like manner \cite{Ashtekar86,Ashtekar87} in space-time rather than space, where the subscript $a$ refers to the Lie algebra of the Lorentz group appearing as the natural gauge group for the above decomposition of $g_{\mu\nu}$. As the definition of $A_{a \mu}$ involves derivatives of the basic variables ${b_\mu}^\kappa$, assuming a Yang-Mills action for the fields $A_{a \mu}$ corresponds to a nested  higher derivative theory with the anticipated fourth-order field equations for ${b_\mu}^\kappa$ \cite{hco231,hco235}. For the coupling of a particle to the gravitational field, Eqs.~(\ref{gravparticlecoupling}) and (\ref{propertimedefg}) have been used, but with an additional factor of the curvature scalar (the Ricci tensor might be a better option; alternatively, one could introduce an additional scalar field as in the Brans-Dicke theory \cite{Jordan,Fierz56,Jordan59,BransDicke61}). The decomposition $g_{\mu\nu} = {b_\mu}^\kappa \, b_{\kappa\nu}$ suggests that Einstein's general relativity might emerge via the pairing of two vector bosons to form a graviton. Similar to the pairing of fermions in the Schwinger model \cite{Schwinger62,LowensteinSwieca71,hcoqft}, which has been used as a toy model for quark confinement in quantum chromodynamics \cite{KogutSusskind75a}, a pairing of bosons would be needed. Whether such a pairing mechanism is essential in order to overcome the unavoidably repulsive interactions arising from vector theories for Lagrangians with first derivatives and linear coupling \cite{Straumann00} is not obvious for the proposed higher derivative theory.

\section{Conclusions}
A Noether-enhanced Legendre transformation has been used to develop an alternative to the Lagrangian and Hamiltonian frameworks for relativistic field theories on Minkowski space. The proposed framework is based on the energy-momentum tensor, which is ideal for coupling subsystems that interact through the exchange of energy and momentum. The field equations are contained in the structure of the divergence of the energy-momentum tensor. By expressing this divergence in terms of auxiliary fields, the number of field equations can be larger than the four conditions arising from the vanishing divergence of the energy-momentum tensor. It may happen that the auxiliary fields are such that we obtain constraints or gauge degrees of freedom rather than independent field equations.

An important question is whether the proposed framework could also suggest an alternative to the canonical and path integral quantization methods associated with the Hamiltonian and Lagrangian frameworks, respectively. For pure Yang-Mills fields, wehave shown that the new framework leads to a covariant canonical quantization procedure. In general, it is natural to expect a formulation based on space-time dependent operators that allow us to have access to the local contributions to energy and momentum and also their transport properties. In other words, we are forced to work in real space and time rather than in Fock space. The most promising approach seems to be algebraic quantum field theory \cite{HaagKastler64,Haag}, where the powerful concept of modular localization \cite{Schroer15} is particularly promising for a rigorous implementation of local energy-momentum conservation.

\appendix

\section{Derivation of energy-momentum tensor}\label{appenymomtens}
We here derive the basic identity (\ref{localconservation}) for the energy-momentum tensor (\ref{enhancedLegendre2}), which reduces to the expression (\ref{enhancedLegendre}) if the Lagrange density ${\cal L}$ does not depend on the second derivatives $\varphi^a_{,\mu ,\nu}$. For ${\cal L} = {\cal L}(\varphi^a, \varphi^a_{,\mu}, \varphi^a_{,\mu ,\nu})$, the chain rule implies
\begin{equation}\label{appLder}
   \frac{\partial{\cal L}}{\partial x^\mu} =
   \varphi^a_{,\mu} \, \frac{\partial{\cal L}}{\partial\varphi^a} +
   \varphi^a_{,\nu,\mu} \, \frac{\partial{\cal L}}{\partial\varphi^a_{,\nu}} +
   \varphi^a_{,\nu,\nu',\mu} \, \frac{\partial{\cal L}}{\partial\varphi^a_{,\nu,\nu'}} .
\end{equation}
The functional derivative of the action $c I = \int {\cal L} \, d^4x$ is given by
\begin{equation}\label{appIfuncder}
   c \, \frac{\delta I}{\delta\varphi^a} =
   \frac{\partial{\cal L}}{\partial\varphi^a} -
   \frac{\partial}{\partial x^\nu} \frac{\partial{\cal L}}{\partial\varphi^a_{,\nu}} +
   \frac{\partial^2}{\partial x^\nu \partial x^{\nu'}}
   \frac{\partial{\cal L}}{\partial\varphi^a_{,\nu,\nu'}} ,
\end{equation}
which implies
\begin{equation}\label{appIfuncder0}
   0 = \varphi^a_{,\mu} \left( c \, \frac{\delta I}{\delta\varphi^a} -
   \frac{\partial{\cal L}}{\partial\varphi^a} +
   \frac{\partial}{\partial x^\nu} \frac{\partial{\cal L}}{\partial\varphi^a_{,\nu}} -
   \frac{\partial^2}{\partial x^\nu \partial x^{\nu'}}
   \frac{\partial{\cal L}}{\partial\varphi^a_{,\nu,\nu'}} \right) .
\end{equation}
By adding Eqs.~(\ref{appLder}) and (\ref{appIfuncder0}), we obtain
\renewcommand{\arraystretch}{1.75}
\begin{eqnarray}
   \frac{\partial{\cal L}}{\partial x^\mu} &-& \frac{\partial}{\partial x^\nu}
   \left( \varphi^a_{,\mu} \frac{\partial{\cal L}}{\partial\varphi^a_{,\nu}} \right)
   + \varphi^a_{,\mu} \frac{\partial^2}{\partial x^\nu \partial x^{\nu'}}
   \frac{\partial{\cal L}}{\partial\varphi^a_{,\nu,\nu'}} \nonumber \\
   &-& \varphi^a_{,\nu,\nu',\mu} \, \frac{\partial{\cal L}}{\partial\varphi^a_{,\nu,\nu'}}
   = c \, \varphi^a_{,\mu} \, \frac{\delta I}{\delta\varphi^a} \rule{0pt}{5ex} .
\label{appclose}
\end{eqnarray}
The left-hand side of Eq.~(\ref{appclose}) can now be identified as the divergence of the energy-momentum tensor (\ref{enhancedLegendre2}) so that we have completed our derivation of the identity (\ref{localconservation}).

%\bibliography{hcopubs}

\end{document}